\documentstyle[prl,aps,psfig,floats]{revtex}

\def\e{{\rm e}}

\begin{document}

\wideabs{
\title{Thermodynamics of the quantum easy-plane antiferromagnet on the
       triangular lattice}
\author{Luca Capriotti}
\address{Scuola Internazionale Superiore di Studi Avanzati,
         via Beirut 2-4, 34013 Trieste, Italy, \\
         and Istituto Nazionale di Fisica della Materia (INFM).}
\author{Alessandro Cuccoli, Valerio Tognetti, Paola Verrucchi}
\address{Dipartimento di Fisica dell'Universit\`a di Firenze,
         Largo E. Fermi~2, I-50125 Firenze, Italy, \\
         and Istituto Nazionale di Fisica della Materia (INFM).}
\author{Ruggero Vaia}
\address{Istituto di Elettronica Quantistica
         del Consiglio Nazionale delle Ricerche,
         via Panciatichi~56/30, I-50127 Firenze, Italy, \\
         and Istituto Nazionale di Fisica della Materia (INFM).}
\date{\today}
\maketitle
\begin{abstract}
The classical XXZ triangular-lattice antiferromagnet (TAF) shows both
an Ising and a BKT transition, related to the chirality and the
in-plane spin components, respectively. In this paper the quantum effects
on the thermodynamic quantities are evaluated by means of the pure-quantum
self-consistent harmonic approximation (PQSCHA), that allows one to deal
with any spin value through classical MC simulations.
We report the internal energy, the specific heat, and the in-plane
correlation length of the quantum XX0 TAF, for $S=1/2$, 1, 5/2.
The quantum transition temperatures turn out to be smaller the smaller
the spin, and agree with the few available theoretical and numerical estimates.
\end{abstract}


} 

A renewed interest has recently focused on triangular antiferromagnets
(TAF)~\cite{Kawacollins}. Indeed, they turned out to describe the magnetic
behavior of several real compounds as, for example, the stacked antiferromagnet
${\rm NaTiO_2}$~\cite{Hir85}, the organic superconductors of the family
$\kappa{-}{\rm (BEDT{-}TTF)_2X}$~\cite{McKenzie97} and the K/Si(111):B
interface~\cite{Weitering97}.

In this paper we investigate the thermodynamic properties of the quantum
XXZ Heisenberg antiferromagnet on the triangular lattice,
defined by the following Hamiltonian
\begin{equation}
 \hat{\cal H}={J\over2}\sum_{{\bf{i}},{\bf{d}}}
 \big( \hat{S}_{\bf{i}}^x \hat{S}_{{\bf{i}}+{\bf{d}}}^x
 + \hat{S}_{\bf{i}}^y \hat{S}_{{\bf{i}}+{\bf{d}}}^y
 +\lambda \, \hat{S}_{\bf{i}}^z \hat{S}_{{\bf{i}}+{\bf{d}}}^z \big)~,
\label{e.xxzmodel}
\end{equation}
where $J$ is the positive (antiferromagnetic) exchange constant, and
$\big(\hat{S}_{\bf{i}}^x,\hat{S}_{\bf{i}}^y,\hat{S}_{\bf{i}}^z\big)$ are
the spin operators sitting on the sites ${\bf{i}}$ of a triangular lattice.
They satisfy ${\cal SU}(2)$ commutation relations and belong to the
spin-$S$ representation $\big|\hat{\bf{S}}_{\bf i}^2\big|=S(S+1)$.
The interaction is restricted to nearest-neighbors and ${\bf{d}}$ runs
over their relative displacements.
The planar character of the system is due to the presence of the anisotropy
$\lambda\in[0,1)$, energetically favoring configurations with the spins
lying in the $xy$ plane (easy-plane).
For $\lambda=0$ the spin components on the $z$ axis do not
appear in the Hamiltonian and the model is known as  XX0 or quantum XY.

The  minimum energy configuration of  the classical counterpart
of the Hamiltonian~(\ref{e.xxzmodel}) for every value of  $\lambda\in[0,1]$
consists of coplanar spins forming
$\pm{2}\pi/3$ angles between nearest-neighbors
and this leads to a $\sqrt{3}\times\sqrt{3}$
periodic N\'eel state. In contrast to the isotropic case, where the
plane in which the $2\pi/3$ structure lies can take any direction in the spin
space, in the XXZ model  such structure must take place in the easy-plane.
As a result, in the planar TAF the frustration causes an
additional discrete two-fold degeneracy of the classical  ground state, which
is due to
 the chirality (or helicity), defined as the sign of rotation of the spins along
the sides of each elementary triangle.
The resulting degeneracy corresponds to the group $SO(2)\times{Z_2}$.
As the Mermin and Wagner theorem only states that the sublattice
magnetization must vanish at any non zero temperature,
long-range order can occur as far as the chirality is concerned,
and an Ising-like phase transition is indeed observed~\cite{CVCT98},
in addition to the usual Berezinskii-Kosterlitz-Thouless
(BKT) critical behavior associated to the rotation symmetry in the $xy$ plane.

In the quantum case the situation is far less clear. In fact, unlike
the antiferromagnet on the square lattice where there is a general
consensus about the ordered nature of the ground state even for $S =
1/2$, in the frustrated cases the lack of exact analytical
results is accompanied by difficulties in applying stochastic numerical
methods, as their reliability is strongly limited by the
well-known sign problem.
Indeed only very recently a systematic size-scaling of the order parameter
and of the spin gap has been performed using a new Quantum
Monte Carlo technique~\cite{CTS99}, confirming the existence
of N\'eel long-range order in the ground state
as also suggested by the symmetry properties of
the first excited states, evidenced by Bernu {\em et al.}~\cite{Bernuetal94}.

An even less clear situation is that concerning the finite temperature
behavior.  In fact an early numerical work~\cite{MaTsu88}, limited to
lattice sizes up to 27 sites, indicated for the $S = 1/2$ XX0 model a
phase diagram similar to the classical one, in contrast with the high
temperature expansion produced by Fujiki and Betts~\cite{FujiBet91}
where no evidence for a phase transition was found. For the XXZ
Hamiltonian, Momoi and Suzuki~\cite{MoSu92}, applying an effective
field theory, conjectured that the chiral phase transition should
persist for every value of $\lambda \in [0,1)$, as in the classical
case, and were able to estimate the transition temperature for
$\lambda = 0$, obtaining a value very close to that found in
Ref.~\cite{MaTsu88}.  Recently Suzuki and Matsubara~\cite{SuMat94}
using a quantum transfer Monte Carlo method to study clusters up to 24
sites, have claimed instead the absence of the chiral order at any
finite temperature for $\lambda \geq 0.6$.

In this context where, at least up to now, quantum numerical
methods cannot give satisfactory answers, the {\em pure-quantum
self-consistent harmonic approximation} (PQSCHA)~\cite{CGTVV95}
can provide an effective instrument to investigate the thermodynamics
of quantum spin systems, as far as their ground state is ordered.
The method is based on the path-integral formulation of
quantum statistical mechanics, and has been successfully
applied recently to a variety of unfrustrated spin models, both
one-~\cite{CTVV92} and two-dimensional~\cite{CTVV96prl,CCTVV98}.

By the PQSCHA the evaluation of thermal averages in the quantum
model can be reduced to the calculation of classical-like averages
over a Boltzmann distribution defined by an effective Hamiltonian,
which contains the contribution of the pure-quantum part of the
fluctuations (approximated within a self consistent harmonic
scheme) in its renormalized interaction parameters,
which are temperature and spin dependent.
As a result one can get accurate results on the
quantum spin system using classical computational methods, like the
transfer-matrix in the one-dimensional case and classical Monte
Carlo simulations in the two-dimensional one.

The first step of the derivation of the effective Hamiltonian for the
easy-plane TAF, is to apply the unitary transformation which defines a
spatially varying coordinate system pointing along the local N\'eel
direction, namely
\begin{equation}
{\cal U}  =  \exp{\left( \frac{2\pi}{3}i \sum_{\bf{i} \in {\rm B}} \hat{S}_{\bf{i}}^z
- \frac{2\pi}{3}i \sum_{\bf{i} \in {\rm C}} \hat{S}_{\bf{i}}^z \right)}~,
\end{equation}
where B and C labels two of the three sublattices.
Unlike in the bipartite lattices where the
corresponding transformation maps the antiferromagnet into a model
with an in-plane ferromagnetic exchange and an antiferromagnetic
coupling along the $z$ axis, thus allowing the demonstration of the Lieb
and Mattis theorem~\cite{LiebMatt62} and computability with standard
quantum Monte Carlo methods, in the triangular case the transformed
Hamiltonian shows an extra current-like term~\cite{LeungRunge93}
which contains the effects of the frustration,
whose form is quite similar to the chiral
order parameter~\cite{CVCT98}, i.e., the physical quantity undergoing
the order-disorder phase transition present in the classical model for
every value of $\lambda<1$.

From now on the derivation follows the same lines already described in
Refs.~\cite{CTVV92,CCTVV98}. A point worth being recalled is the use of
the Villain transformation~\cite{Villain74} in order to represent the
spin operators in terms of
canonically conjugated variables, which is a necessary step in the
derivation of the effective Hamiltonian. As it is well known, this
spin-boson transformation preserves the commutation rules but neglects
the so called kinematic interaction due to the limited spectrum of
$S^z$, thus giving a better description when the system has a good
easy-plane character and the spin states with large fluctuations of
$S^z$ are less relevant to the thermodynamics. In the square
lattice case~\cite{CCTVV98} such approximation scheme turns out to be
reliable up to some value of
$\lambda_{\rm M}<1$ ($\lambda_{\rm M} = 0.58$ in the
extreme quantum case $S = 1/2$), when the mapping with the Villain
transformation breaks down and a different spin-boson transformation
is needed. However, it provides accurate results for the critical
temperatures even for $\lambda = 0.5$; a similar behavior is also
found in the case of the quantum TAF. Finally we remind that Weyl
ordering, which is inherent to the PQSCHA, naturally leads to define
an effective classical spin length as $\tilde{S}=S+1/2$ and thus to set the
natural energy scale $\epsilon =J\tilde{S}^2$. Therefore in the
following we use the reduced temperature, $t=k_{B}T/J\tilde{S}^2$.

In the case of the easy-plane TAF the effective Hamiltonian
has the form ${\cal H}_{\rm eff}=\overline{\cal H}+G(t)$, where $G(t)$
is an additive uniform
term, formally identical to that obtained for the square lattice,
which is unessential in the calculation of the thermal averages, while:
\begin{equation}
\overline{\cal H}={\epsilon \over 2}j_{\rm eff}\sum_{{\bf{i}},{\bf{d}}}
 \big( s_{\bf{i}}^x s_{{\bf{i}}+{\bf{d}}}^x
 + s_{\bf{i}}^y s_{{\bf{i}}+{\bf{d}}}^y
 +\lambda_{\rm eff} \, s_{\bf{i}}^z s_{{\bf{i}}+{\bf{d}}}^z \big)~,
\end{equation}
where $\big(s_{\bf{i}}^x,s_{\bf{i}}^y,s_{\bf{i}}^z\big)$ are
unit vectors, i.e., classical spins. Within the PQSCHA quantum
effects are embodied in the temperature and spin dependence of the
renormalized
\begin{eqnarray}
 j_{\rm{eff}}(t,S,\lambda)&=& (1-\frac{1}{2}{D_\perp})^2
 ~e^{-\frac{1}{2}  {\cal{D}}_\|}~,
\label{e.VTjeff}
\\
 \lambda_{\rm{eff}}(t,S,\lambda)&=& \lambda
 ~(1-\frac{1}{2} D_\perp)^{-1}
 ~e^{\frac{1}{2}  {\cal{D}}_\|}~,
\label{e.VTleff}
\end{eqnarray}
with
\begin{eqnarray}
 D_\perp&=&(2\tilde S N)^{-1} {\sum}_{\bf{k}}~ {b_{\bf{k}}\over
a_{\bf{k}}}
 {\cal{L}}(f_{\bf{k}})~,
\label{e.VTDperp}
\\
 {\cal{D}}_\|&=&(\tilde S N)^{-1} {\sum}_{\bf{k}}~ (1-\gamma_{\bf{k}})
 {a_{\bf{k}}\over b_{\bf{k}}}
 {\cal{L}}(f_{\bf{k}})~,
\label{e.VTDpar}
\end{eqnarray}
where
\begin{eqnarray}
 a^2_{\bf{k}}&=&\frac{z}{2}(1-{\textstyle{1\over2}}{D_\perp})
 ~\e^{-{\textstyle{1\over2}}{{\cal{D}}_\|}}
 ~(1+2\lambda_{\rm{eff}}\,\gamma_{\bf{k}})~,
\\
 b^2_{\bf{k}}&=&\frac{z}{2}j_{\rm eff}(1-\gamma_{\bf{k}})~,
\end{eqnarray}
$f_{\bf{k}}=a_{\bf{k}}{b_{\bf{k}}}/(2\tilde S{t})$~,
${\cal{L}}(x)=\coth x-x^{-1}$ is the Langevin function,
$\gamma_{\bf{k}}=z^{-1}{\sum}_{\bf{d}}\cos({\bf{k}}{\cdot}{\bf{d}})$~,
and ${\bf{k}}$ is a wavevector varying in the first
Brillouin zone. $D_\perp(S,\lambda,t)$ and
${\cal{D}}_\|(S,\lambda,t)$ represent the pure-quantum square
fluctuations of the out-of-plane and in-plane components of
the spins respectively. They are decreasing functions of
temperature and spin, vanishing for $
t\rightarrow\infty$ and $S\rightarrow\infty$, i.e., when the
quantum part of the fluctuations is negligible with respect to
the classical one.

From the above equations, we can infer that the PQSCHA approach is
valid under the condition that second order terms in $(D_\perp)^2$ can be
neglected. One can take the criterion that the renormalization effects of
quantum fluctuations must not reduce much more than, say,  $50\%$ the
effective exchange integral.
Such a strong renormalization only occurs for $S=1/2$ and
$t\lesssim 0.2$, while for higher spin values the PQSCHA is reliable
at any temperature.

In the XX0 model, $\lambda_{\rm eff} =\lambda=0$ and all the information
about the quantum system is hence contained in the renormalization of the
energy scale. In this case the critical properties of the quantum system at
a temperature $t$ are essentially those of its classical counterpart at the
effective temperature $t_{\rm eff} = t/j_{\rm eff}(t,S)$, and we have used
the results of classical Monte Carlo (MC) simulations recently obtained for
lattice sizes up to $N=120 \times 120$~\cite{CVCT98} to calculate the
corresponding quantum observables.

\begin{figure}
\centerline{\psfig{bbllx=60pt,bblly=250pt,bburx=512pt,bbury=640pt,%
figure=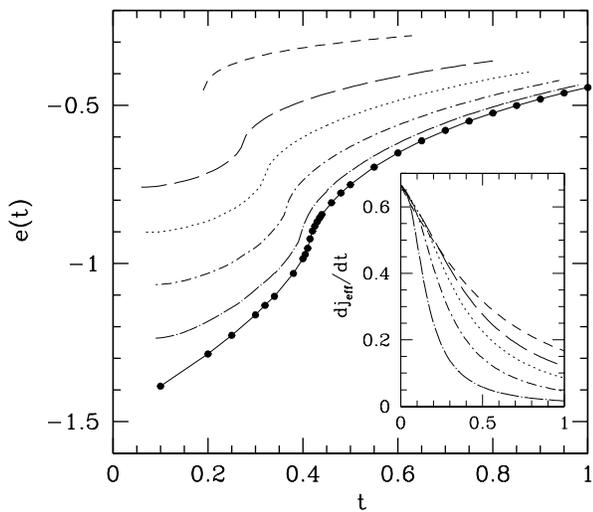,width=80mm,angle=0}}
\caption{Temperature and spin dependence of the internal energy per spin
  for $\lambda=0$ and (from the top curve) $S = 1/2$,$1$,$3/2$,$5/2$,$5$
  and $\infty$. Solid circles are classical MC data. In the inset the
  derivative of the effective exchange constant is plotted vs. $t$
  with the same convention on the lines.}
\label{f.erg}
\end{figure}

Indicating the classical averages with the
effective Hamiltonian as ${\langle \dots \rangle} _{\rm eff}$,
the internal energy per spin can be calculated as
\begin{equation}
 e(t,S,\lambda)=\frac {\langle {\hat {\cal H} } \rangle}{N\epsilon}
 = {\langle \overline{\cal H} \rangle} _{\rm eff}
 + {z\over2}\,\lambda\, {\cal D}_\perp~,
\label{int.erg}
\end{equation}
where ${\cal D}_\perp$ can be expressed as $D_\perp$, Eq.~(\ref{e.VTDperp}),
with an extra factor $\gamma_{\bf{k}}$ in the summand.
For $\lambda =0$ the above equation reduces to
\begin{equation}
 e(t,S)=j_{\rm eff}(t)\,e_{\rm cl}(t_{\rm eff}),
\label{e.erg0}
\end{equation}
being $e_{\rm cl}(t)$ the internal energy per
spin of the corresponding classical system. In Fig.~\ref{f.erg} $e(t,S)$
is plotted for various values of the spin in the range of temperatures
where the PQSCHA is expected to give reliable results.

The energy curves flatten and increase with decreasing $S$ due to the
increased quantum fluctuations. As said before the $S=1/2$ curve is
reported only in the valid temperature range. As a matter of fact the
extrapolation to lowest temperatures gives the self-consistent spin-wave
ground-state energy. The difference from the most refined
estimates~\cite{LeungRunge93} can be mainly attributed to $1/S^2$
constant contributions coming from the Villain
transformation~\cite{NishimoriM85} and to the use of the low coupling
approximation (LCA)~\cite{CGTVV95}. This term is not significant
for $S\geq1$.

\begin{figure}
\centerline{\psfig{bbllx=75pt,bblly=250pt,bburx=510pt,bbury=570pt,%
figure=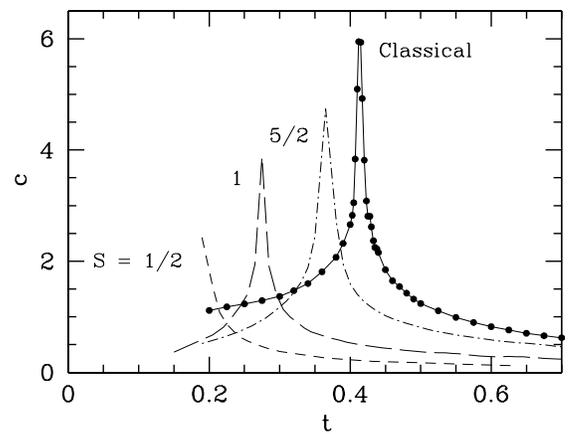,width=80mm,angle=0}}
\caption{Temperature and spin dependence of the specific heat
in the XX0 model. Circles are the classical MC data obtained
from the mean squared fluctuations of the energy while the solid line
is the  numerical derivative of the energy curve. }
\label{f.sph}
\end{figure}

Consistently with this picture the finite size ($N=120\times 120$)
peak of the specific heat (Fig.~\ref{f.sph}), obtained by numerical
derivation of the internal energy, moves towards lower temperatures
and decreases in height as $S$ decreases. However, since the quantum
renormalizations are essentially size-independent, classical scaling with
size~\cite{CVCT98} is conserved and a logarithmic divergence of the
specific heat, connected with the Ising-like chirality
phase-transition, is therefore expected in the thermodynamic limit.
By direct derivation of Eq.~(\ref{e.erg0}) it is easily seen that,
in order for the quantum specific heat to vanish in
the zero temperature limit, within our approximation we must have
$d\,j_{\rm eff}/dt\to|e_{\rm cl}(0)|^{-1}$ as $t\to0$
a condition which is fulfilled for every value of $S$, as can be verified
analytically from the explicit expressions of the renormalization parameters;
this is shown in the inset of Fig.~\ref{f.erg}.

\begin{figure}
\centerline{\psfig{bbllx=75pt,bblly=250pt,bburx=510pt,bbury=560pt,%
figure=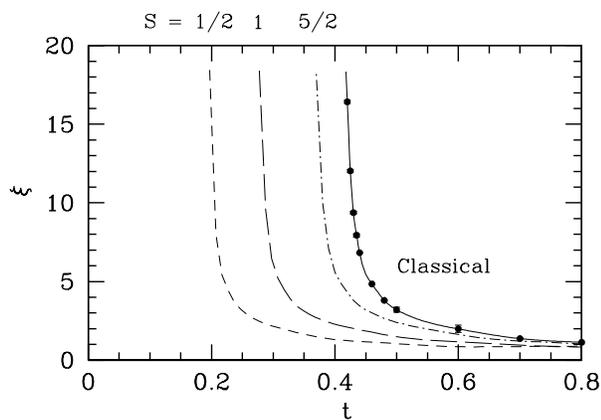,width=80mm,angle=0}}
\caption{Temperature and spin dependence of the magnetic correlation length
in the XX0 model. Circles and solid line are classical MC data.}
\label{f.xi}
\end{figure}

Most papers in literature do mainly concern with the chiral order-disorder
transition, while the XXZ TAF also supports another kind of
phase transition.
In fact, the classical system~\cite{CVCT98} displays as well a BKT
critical behavior. For this reason we have calculated
the magnetic correlation length which governs the decay of
the in-plane correlation functions in the
high-temperature phase, whose expression within the PQSCHA reads
\begin{equation}
 \langle \hat{S}^x_{\bf i}\hat{S}^x_{{\bf i}+{\bf r }}+
 \hat{S}^y_{\bf i}\hat{S}^y_{{\bf i}+{\bf r }} \rangle
 = G({\bf r})\, \langle s^x_{\bf i} s^x_{{\bf i}+{\bf r }}+
 s^y_{\bf i} s^y_{{\bf i}+{\bf r }} \rangle_{\rm{eff}}~,
\end{equation}
where ${\bf i}$ and ${\bf i{+}r}$ belong to the same sublattice and
$G({\bf r})$ is bounded and essentially constant for large ${\bf r}$.
As a consequence  the asymptotic behavior of the correlation functions in
the critical region is the same of the effective classical spin system.
In particular, for $\lambda=0$, the correlation length can be simply found
as $\xi(t)=\xi_{\rm cl}(t_{\rm eff})$. The result is reported in
Fig.~\ref{f.xi}: as expected in a BKT transition, it displays
a divergence at a temperature $t_{\rm{BKT}}$ which decreases
with decreasing $S$, as a result of enhanced quantum fluctuations.

Within the PQSCHA, the quantum renormalizations cannot modify the critical
behavior of the effective classical system~\cite{note},
 so that both the chirality and
the BKT critical temperatures can be connected to their classical
counterparts by the self-consistent relation
\begin{equation}
\frac{t_{\rm crit}(S,\lambda)}{t^{\rm cl}_{\rm crit}
\big(  \lambda_{\rm eff}(t_{\rm crit},S,\lambda) \big)}
=j_{\rm eff}(t_{\rm crit},S,\lambda)~,
\label{e.sctc}
\end{equation}
which can be solved numerically.
The obtained critical temperatures for $\lambda=0$ and $\lambda=0.5$ are
reported for various values of the spin in Table~\ref{t.tcrit}.
In the $S=1/2$ case, although our theory begins to become unreliable
when $t\lesssim{0.2}$, we notice that for $\lambda=0$ the
extrapolated value for the chiral critical temperature, $t_{\rm c}=0.193$,
agrees remarkably well with those obtained by the size scaling on
the QMC~\cite{MaTsu88} data ($t_{\rm c}=0.195(1)$) and the
effective field theory of Ref.~\cite{MoSu92} ($t_{\rm c}\simeq 0.20$).

\medskip

L.~C. acknowledges S. Sorella for continuous and fruitful discussions.

\squeezetable

\begin{table}
\caption{Chiral and BKT critical temperatures for $\lambda=0$ and $\lambda=0.5$
and for some values of the spin length  $S$.
The classical values are taken from Ref.~\protect\cite{CVCT98}.
The reported errors only represent the statistical uncertainty of the MC data.}
\begin{tabular}{ccccccc}
 $S$                &  1/2      &  1       &  3/2      &  5/2       &   5        & $\infty $ \\
\tableline\\
 $t_{\rm c}(S,0)$     & 0.193(2)  & 0.273(3) & 0.319(3)  &  0.364(4)  &  0.396(5)  &   0.412(5)  \\
 $t_{\rm BKT}(S,0)$   & 0.1875(5) & 0.265(1) & 0.310(1)  &  0.352(1)  &  0.386(1)  &   0.402(2)  \\
\\
 $t_{\rm c}(S,0.5)$   & 0.185(4)  & 0.267(4) & 0.312(5)  &  0.355(5)  &  0.385(5)  &   0.400(5)  \\
 $t_{\rm BKT}(S,0.5)$ & 0.180(1)  & 0.260(1) & 0.304(2)  &  0.346(2)  &  0.376(2)  &   0.391(2)  \\
\end{tabular}
\label{t.tcrit}
\end{table}


\begin{thebibliography}{200}
\bibitem{note}
This is also true with respect to the chiral critical behavior
being the derivatives of $j_{\rm eff}$ and $\lambda_{\rm eff}$ bounded.
\end{thebibliography}
\end{document}